\documentclass{article}

\PassOptionsToPackage{numbers, compress}{natbib}

    \usepackage[preprint]{neurips_2023}

\usepackage[utf8]{inputenc} 
\usepackage[T1]{fontenc}    
\usepackage[hidelinks]{hyperref}       
\usepackage{url}            
\usepackage{booktabs}       
\usepackage{amsfonts}       
\usepackage{nicefrac}       
\usepackage{xcolor}         
\usepackage{tikz}
\usepackage{pifont}
\usepackage{amsmath,amssymb}
\usepackage{multirow}
\usepackage{fontawesome5}
\usepackage{soul}
\usepackage[capitalize]{cleveref}
\usepackage{enumitem}

\newcommand{\cmark}{\ding{51}}%
\newcommand{\xmark}{\ding{55}}%

\definecolor{green}{rgb}{0.35, 0.90, 0.63}

\newcommand{\greencheck}{{\color{green}\cmark}}
\newcommand{\redcross}{{\color{red}\xmark}}

\newcommand{\model}{\textsc{Tango}}

\title{Text-to-Audio Generation using Instruction-Tuned LLM and Latent Diffusion Model}

\let\realcite\cite
\renewcommand{\cite}[1]{\ifx.#1.\hl{[?]}\else\realcite{#1}\fi}
\let\realcitet\citet
\renewcommand{\citet}[1]{\ifx.#1.\hl{who at al. [?]}\else\realcitet{#1}\fi}

\author{%
  Deepanway Ghosal\textsuperscript{\ddag}, Navonil Majumder\textsuperscript{\ddag}, Ambuj Mehrish\textsuperscript{\ddag}, Soujanya Poria\textsuperscript{\ddag} \\
  \textsuperscript{\ddag} DeCLaRe Lab, Singapore University of Technology and Design, Singapore \\
  \texttt{deepanway\_ghosal@mymail.sutd.edu.sg} \\
  \texttt{\{navonil\_majumder,ambuj\_mehrish,sporia\}@sutd.edu.sg} \\
}

\begin{document}
\maketitle
\begin{minipage}[t]{\linewidth}
  \begin{center}
    \includegraphics[width=0.5\linewidth]{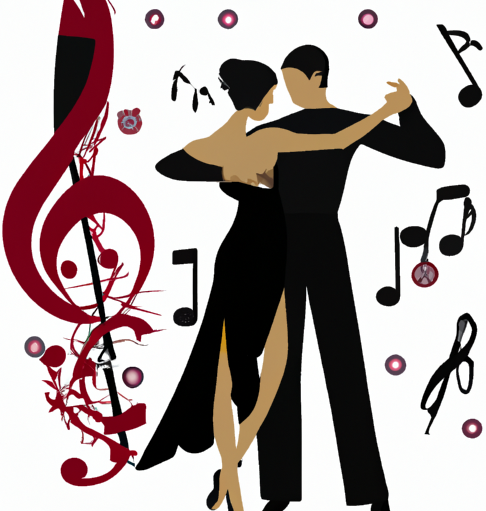}
  \end{center}
\end{minipage}
\vspace{0.3cm}
\begin{minipage}[t]{\linewidth}
  \centering
  \faGithub: \url{https://github.com/declare-lab/tango} \\
  \faGlobe : \url{https://tango-web.github.io/}
\end{minipage}

\begin{abstract}
The immense scale of the recent large language models (LLM) allows many interesting properties, such as, instruction- and chain-of-thought-based fine-tuning, that has significantly improved zero- and few-shot performance in many natural language processing (NLP) tasks. Inspired by such successes, we adopt such an instruction-tuned LLM \textsc{Flan-T5} as the text encoder for text-to-audio (TTA) generation---a task where the goal is to generate an audio from its textual description. The prior works on TTA either pre-trained a joint text-audio encoder or used a non-instruction-tuned model, such as, \texttt{T5}. Consequently, our latent diffusion model (LDM)-based approach (\model{}) outperforms the state-of-the-art AudioLDM on most metrics and stays comparable on the rest on AudioCaps test set, despite training the LDM on a 63 times smaller dataset and keeping the text encoder frozen. This improvement might also be attributed to the adoption of audio pressure level-based sound mixing for the training set augmentation, whereas the prior methods take a random mix. 
\end{abstract}

\section{Introduction}

Following the success of automatic text-to-image (TTI) generation~\cite{Ramesh2021ZeroShotTG,Ramesh2022HierarchicalTI,rombach2022high}, many researchers have also succeeded in text-to-audio (TTA) generation~\cite{Kreuk2022AudioGenTG,Liu2023AudioLDMTG,yang2022diffsound} by employing similar techniques as the former. Such models may have strong potential use cases in the media production where the creators are always looking for novel sounds that fit their creations. This could be especially useful in prototyping or small-scale projects where producing the exact sound could be infeasible. Beyond this, these techniques also pave the path toward general-purpose multimodal AI that can simultaneously recognize and generate multiple modalities. 

To this end, the existing works use a large text encoder, such as, \texttt{RoBERTa}~\cite{Liu2019RoBERTaAR} and \texttt{T5}~\cite{2020t5}, to encode the textual description of the audio to be generated. Subsequently, a large transformer decoder or a diffusion model generates the audio prior, which is subsequently decoded by a pre-trained VAE, followed by a vocoder. We instead assume that replacing the text encoder with an instruction-tuned large language model (LLM) would improve text understanding and overall audio generation without any fine-tuning, due to its recently discovered gradient-descent mimicking property~\cite{Dai2022WhyCG}. To augment training samples, the existing methods take a randomly generated combination of audio pairs, along with the concatenation of their descriptions. Such a mixture does not account for the overall pressure level of the source audios, potentially leading to a louder audio overwhelming the quieter one. Thus, we employ a pressure level-based mixing method, as suggested by \citet{DBLP:journals/corr/abs-1711-10282}.

Our model (\model{})~\footnote{The acronym \model{} stands for \texttt{\textbf{T}ext-to-\textbf{A}udio using i\textbf{N}struction \textbf{G}uided diffusi\textbf{O}n} and was suggested by \texttt{ChatGPT}. The word \model{} is often associated with music~\cite{wiki:tango-music} and dance~\cite{wiki:tango}. According to Wikipedia~\cite{wiki:tango}, ``Tango is a partner dance and social dance that originated in the 1880s along the Río de la Plata, the natural border between Argentina and Uruguay.'' The image above resembles the \model{} dance form and was generated by prompting \texttt{Dalle-V2} with \texttt{``A couple dancing tango with musical notes in the background''}} is inspired by latent diffusion model (LDM)~\cite{rombach2022high} and AudioLDM~\cite{Liu2023AudioLDMTG} models. However, instead of using CLAP-based embeddings, we used a large language model (LLM) due to its powerful representational ability and fine-tuning mechanism, which can help learn complex concepts in the textual description. Our experimental results show that using an LLM greatly improves text-to-audio generation and outperforms state-of-the-art models, even when using a significantly smaller dataset. In the image generation literature, the effects of LLM has been studied before by \citet{saharia2022photorealistic}. However, they considered T5 as the text encoder which is not pre-trained on instruction-based datasets. \textsc{Flan-T5}~\citep{https://doi.org/10.48550/arxiv.2210.11416} is initialized with a T5 checkpoint and fine-tuned on a dataset of 1.8K NLP tasks in terms of instructions and chain-of-thought reasoning. By leveraging instruction-based tuning, \textsc{Flan-T5} has achieved state-of-the-art performance on several NLP tasks, matching the performance of LLMs with billions of parameters.

In \cref{sec:experiments}, we empirically show that \model {} outperforms AudioLDM and other baseline approaches on most of the metrics on AudioCaps test set under both objective and subjective evaluations, despite training the LDM on a $63$ times smaller dataset. We believe that if \model{} is trained on a larger dataset such as AudioSet (as \citet{Liu2023AudioLDMTG} did), it would be able to provide even better results and improve its ability to recognize a wider range of sounds.

The overall contribution of this paper is threefold:
\begin{enumerate}[leftmargin=*, wide, labelwidth=0pt, labelindent=0pt]
    \item We do not use any joint text-audio encoder---such as CLAP---for guidance. \citet{Liu2023AudioLDMTG} claim that CLAP-based audio guidance is necessary during training for better performance. We instead use a frozen instruction-tuned pre-trained LLM \textsc{Flan-T5} with strong text representation capacity for text guidance in both training and inference.
    \item AudioLDM needed to fine-tune RoBERTa~\cite{Liu2019RoBERTaAR} text encoder to pre-train CLAP. We, however, keep \textsc{Flan-T5} text encoder frozen during LDM training. Thus, we find that LDM itself is capable of learning text-to-audio concept mapping and composition from a 63 times smaller training set, as compared to AudioLDM, given an instruction-tuned LLM.
    \item To mix audio pairs for data augmentation, inspired by \citet{DBLP:journals/corr/abs-1711-10282}, we consider the pressure levels of the audio pairs, instead of taking a random combination as the prior works like AudioLDM. This ensures good representations of both source audios in the fused audio.
\end{enumerate}

\section{Method}

\begin{figure}
    \centering
    \includegraphics[width=\textwidth]{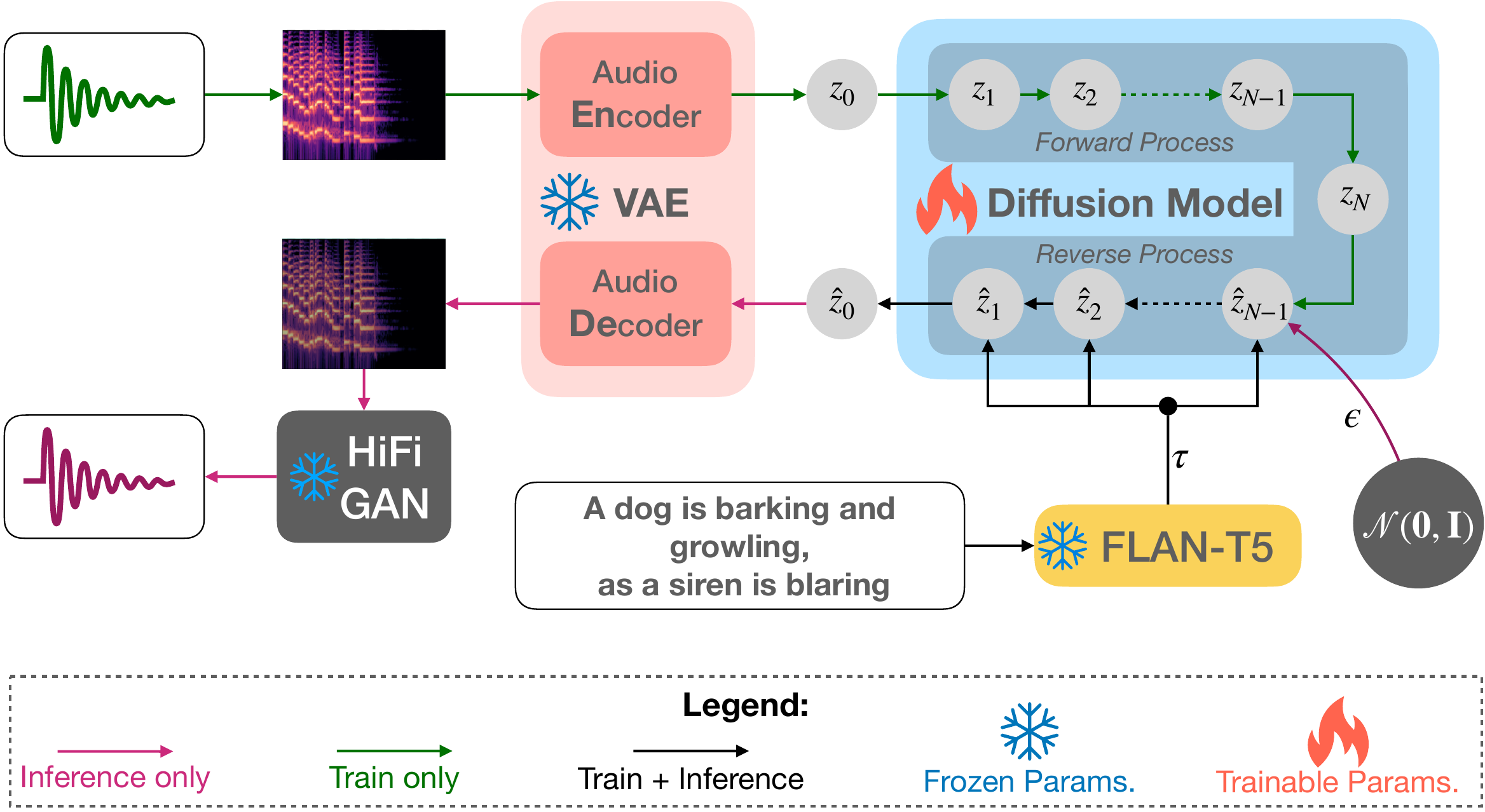}
    \caption{Overall architecture of \model{}.}
    \label{fig:model}
\end{figure}

\model{}, as depicted in \cref{fig:model}, has three major components: i) textual-prompt encoder, ii) latent diffusion model (LDM), and iii) mel-spectogram/audio VAE. The textual-prompt encoder encodes the input description of the audio. Subsequently, the textual representation is used to construct a latent representation of the audio or audio prior from standard Gaussian noise, using reverse diffusion. Thereafter the decoder of the mel-spectogram VAE constructs a mel-spectogram from the latent audio representation. This mel-spectogram is fed to a vocoder to generate the final audio.

\subsection{Textual-Prompt Encoder}
\label{sec:TE}

We use the pre-trained LLM \textsc{Flan-T5-Large} (780M)~\cite{https://doi.org/10.48550/arxiv.2210.11416} as the text encoder ($E_{text}$) to obtain text encoding $\tau\in \mathbb{R}^{L\times d_{text}}$, where $L$ and $d_{text}$ are the token count and token-embedding size, respectively. Due to the pre-training of \textsc{Flan-T5} models on a large-scale chain-of-thought- (CoT) and instruction-based dataset, \citet{Dai2022WhyCG} posit that they are able to learn a new task very well from the in-context information by mimicking gradient descent through attention weights. This property is missing in the older large models, such as \texttt{RoBERTa}~\cite{Liu2019RoBERTaAR} (used by \citet{Liu2023AudioLDMTG}) and \texttt{T5}~\cite{2020t5} (used by \citet{Kreuk2022AudioGenTG}). Considering each input sample a distinct task, it might be reasonable to assume that the gradient-descent mimicking property could be pivotal in learning the mapping between textual and acoustic concepts without fine-tuning the text encoder. The richer pre-training may also allow the encoder to better emphasize the key details with less noise and enriched context. This again may lead to the better transformation of the relevant textual concepts into their acoustics counterparts. Consequently, we keep the text encoder frozen, assuming the subsequent reverse diffusion process (see \cref{sec:LDM}) would be able to learn the inter-modality mapping well for audio prior to construction. We also suspect that fine-tuning $E_{text}$ may degrade its in-context learning ability due to gradients from the audio modality that is out of distribution to the pre-training dataset. This is in contrast with \citet{Liu2023AudioLDMTG} that fine-tunes the pre-trained text encoder as a part of the text-audio joint-representation learning (CLAP) to allow audio prior reconstruction from text. In \cref{sec:experiments}, we empirically show that such joint-representation learning may not be necessary for text-to-audio transformation.

\subsection{Latent Diffusion Model for Text-Guided Generation}
\label{sec:LDM}

The latent diffusion model (LDM)~\cite{rombach2022high} is adapted from \citet{Liu2023AudioLDMTG}, with the goal to construct the audio prior $z_0$ (see \cref{sec:VAE}) with the guidance of text encoding $\tau$. This essentially reduces to approximating the true prior $q(z_0|\tau)$ with parameterized $p_\theta(z_0|\tau)$.

LDM can achieve the above through forward and reverse diffusion processes. The forward diffusion is a Markov chain of Gaussian distributions with scheduled noise parameters $0 < \beta_1 < \beta_2 < \cdots < \beta_N < 1$ to sample noisier versions of $z_0$:
\begin{flalign}
q(z_n|z_{n-1}) &= \mathcal{N}(\sqrt{1-\beta_n} z_{n-1}, \beta_n \mathbf{I}),  \label{eq:forward_diff}\\
q(z_n|z_0) &= \mathcal{N}(\sqrt{\overline\alpha_n} z_0, (1-\overline\alpha_n)\mathbf{I}) \label{eq:quick_forward_diff},
\end{flalign}
where $N$ is the number of forward diffusion steps, $\alpha_n = 1 - \beta_n$, and $\overline\alpha_n = \prod_{i=1}^n \alpha_n$. \citet{Song2020DenoisingDI} show that \cref{eq:quick_forward_diff} conveniently follows from \cref{eq:forward_diff} through reparametrization trick that allows direct sampling of any $z_n$ from $z_0$ via a non-Markovian process: 
\begin{equation}
z_n = \sqrt{\overline\alpha_n} z_0 + (1-\overline\alpha_n)\epsilon, \label{eq:z_n_sampling}
\end{equation}
where the noise term $\epsilon\sim \mathcal{N}(\mathbf{0}, \mathbf{I})$. The final step of the forward process yields $z_N\sim \mathcal{N}(\mathbf{0}, \mathbf{I})$.

The reverse process denoises and reconstructs $z_0$ through text-guided noise estimation ($\hat{\epsilon}_\theta$) using loss
\begin{align}
    \mathcal{L}_{DM} = \sum_{n=1}^N\gamma_n \mathbb{E}_{ \epsilon_n\sim \mathcal{N}(\mathbf{0}, \mathbf{I}), z_0} || \epsilon_n - \hat\epsilon_\theta^{(n)}(z_n, \tau) ||_2^2,
\end{align}
where $z_n$ is sampled from \cref{eq:z_n_sampling} using standard normal noise $\epsilon_n$, $\tau$ is the text encoding (see \cref{sec:TE}) for guidance, and $\gamma_n$ is the weight of reverse step $n$~\cite{hang2023efficient}, taken to be a measure of signal-to-noise ratio (SNR) in terms of $\alpha_{1:N}$. The estimated noise is used to reconstruct $z_0$:
\begin{flalign}
    p_\theta(z_{0:N}|\tau) &= p(z_N) \prod_{n=1}^N p_\theta(z_{n-1}|z_n, \tau), \\
    p_\theta(z_{n-1}|z_n, \tau) &= \mathcal{N}(\mu^{(n)}_\theta(z_n, \tau), \Tilde{\beta}^{(n)}),\\
    \mu_\theta^{(n)}(z_n, \tau) &= \frac{1}{\sqrt{\alpha_n}}[z_n - \frac{1 - \alpha_n}{\sqrt{1 - \overline\alpha_n}}\hat\epsilon_\theta^{(n)}(z_n, \tau)],\\
    \Tilde{\beta}^{(n)} &=\frac{1 - \bar{\alpha}_{n-1}}{1 - \bar{\alpha}_n} \beta_n.
\end{flalign}
The noise estimation $\hat\epsilon_\theta$ is parameterized with U-Net~\cite{10.1007/978-3-319-24574-4_28} with a cross-attention component to include the text guidance $\tau$. In contrast, AudioLDM~\cite{Liu2023AudioLDMTG} uses audio as the guidance during training. During inference, they switch back to text guidance, as this is facilitated by pre-trained joint text-audio embedding (CLAP). We did not find audio-guided training and pre-training CLAP to be necessary, as argued in \cref{sec:TE}.

\subsection{Augmentation}
\label{sec:augmentation}
Many text-to-image~\cite{popov2021diffusion} and text-to-audio~\cite{Kreuk2022AudioGenTG} works have shown the efficacy of training with fusion-based augmented samples to improve cross-modal concept-composition abilities of the diffusion network. Therefore, we synthesize additional text-audio pairs by superimposing existing audio pairs on each other and concatenating their captions.

Unlike \citet{Liu2023AudioLDMTG} and \citet{Kreuk2022AudioGenTG}, to mix audio pairs, we do not take a random combination of them. Following \citet{DBLP:journals/corr/abs-1711-10282}, we instead consider the human auditory perception for fusion. Specifically, the audio pressure level $G$ is taken into account to ensure that a sample with high pressure level do not overwhelm the sample with low pressure level. The weight of an audio sample ($x_1$) is calculated as a relative pressure level (see \cref{fig:relative_gain_dist} in the appendix for its distribution)
\begin{flalign}
    p = (1 + 10^\frac{G_1 - G_2}{20})^{-1}, \label{eq:gain}
\end{flalign}
where $G_1$ and $G_2$ are pressure levels of two audio samples $x_1$ and $x_2$, respectively. This ensures good representation of both audio samples, post mixing.

Furthermore, as pointed out by \citet{DBLP:journals/corr/abs-1711-10282}, the energy of a sound wave is proportional to the square of its amplitude. Thus, we mix $x_1$ and $x_2$ as
\begin{flalign}
    \text{mix}(x_1, x_2) = \frac{p x_1 + (1-p) x_2}{\sqrt{p^2 + (1-p)^2}}.
\end{flalign}

\subsection{Classifier-Free Guidance}

To guide the reverse diffusion process to reconstruct the audio prior $z_0$, we employ a classifier-free guidance~\cite{Ho2022ClassifierFreeDG} of text input $\tau$. During inference, a guidance scale $w$ controls the contribution of text guidance to the noise estimation $\hat\epsilon_\theta$, with respect to unguided estimation, where empty text is passed:
\begin{flalign}
    \hat\epsilon_\theta^{(n)}(z_n, \tau) = w \epsilon_\theta^{(n)}(z_n, \tau) + (1 - w) \epsilon_\theta^{(n)}(z_n).
\end{flalign}

We also trained a model for which the text guidance was randomly dropped for 10\% of the samples during training. We found this model to perform equivalently to a model for which text guidance was always used for all samples.

\subsection{Audio VAE and Vocoder}
\label{sec:VAE}

Audio variational auto-encoder (VAE)~\cite{Kingma2013AutoEncodingVB} compresses the mel-spectogram of an audio sample, $m\in \mathbb{R}^{T\times F}$, into an audio prior $z_0\in \mathbb{R}^{C\times T/r\times F/r}$, where $C$, $T$, $F$, $r$ are the number of channels, number of time-slots, number of frequency-slots, and compression level, respectively. The LDM (see \cref{sec:LDM}) reconstructs the audio prior $\hat z_0$ using input-text guidance $\tau$. 
The encoder and decoder are composed of ResUNet blocks~\cite{Kong2021DecouplingMA} and are trained by maximizing evidence lower-bound (ELBO)~\cite{Kingma2013AutoEncodingVB} and minimizing adversarial loss~\cite{Isola2016ImagetoImageTW}.
We adopt the checkpoint of audio VAE provided by \citet{Liu2023AudioLDMTG}. Thus, we use their best reported setting, where $C$ and $r$ are set to $8$ and $4$, respectively.

As a vocoder to turn the audio-VAE decoder-generated mel-spectogram into an audio, we also use HiFi-GAN~\cite{kong2020hifi} as \citet{Liu2023AudioLDMTG}.

\section{Experiments}
\label{sec:experiments}

\subsection{Datasets and Training}
\paragraph{Text-to-Audio Generation.}
We perform our main text-to-audio generation experiments on the AudioCaps dataset~\cite{kim2019audiocaps}. The dataset contains 45,438 audio clips paired with human-written captions for training. The validation set contains 2,240 instances. The audio clips are ten seconds long and were collected from YouTube videos. The clips were originally crowd-sourced as part of the significantly larger AudioSet dataset~\cite{gemmeke2017audio} for the audio classification task.

We train our LDM using only the paired (text, audio) instances from the AudioCaps dataset. We use the AudioCaps test set as our evaluation data. The test set contains five human-written captions for each audio clip. We use one caption for each clip chosen at random following ~\citet{Liu2023AudioLDMTG} for consistent evaluation with their work. The randomly chosen caption is used as the text prompt, using which we generate the audio signal from our model.

\paragraph{Audio VAE and Vocoder.}
We use the audio VAE model from ~\citet{Liu2023AudioLDMTG}. This VAE network was trained on the AudioSet, AudioCaps, Freesound\footnote{https://freesound.org/}, and BBC Sound Effect Library\footnote{https://sound-effects.bbcrewind.co.uk} (SFX) datasets. Longer audio clips in Freesound and BBC SFX were truncated to the first thirty seconds and then segmented into three parts of ten seconds each. All audio clips were resampled in 16KHz frequency for training the VAE network. We used a compression level of 4 with 8 latent channels for the VAE network. 

We also use the vocoder from ~\citet{Liu2023AudioLDMTG} for audio waveform generation from the mel spectrogram generated by the VAE decoder. The vocoder is a HiFi-GAN~\cite{kong2020hifi} network trained on the AudioSet dataset. All audio clips were resampled at 16KHz for training the vocoder network.

\paragraph{Model, Hyperparameters, and Training Details}
We freeze the \textsc{Flan-T5-Large} text encoder in \model{} and only train the parameters of the latent diffusion model. The diffusion model is based on the Stable Diffusion U-Net architecture~\cite{rombach2022high,10.1007/978-3-319-24574-4_28} and has a total of 866M parameters. We use 8 channels and a cross-attention dimension of 1024 in the U-Net model.

We use the AdamW optimizer~\cite{loshchilov2017decoupled} with a learning rate of 3e-5 and a linear learning rate scheduler for training. We train the model for 40 epochs on the AudioCaps dataset and report results for the checkpoint with the best validation loss, which we obtained at epoch 39. We use four A6000 GPUs for training \model{}, where it takes a total of 52 hours to train 40 epochs, with validation at the end of every epoch. We use a per GPU batch size of 3 (2 original + 1 augmented instance) with 4 gradient accumulation steps. The effective batch size for training is $3$ (instance) $* 4$ (accumulation) $* 4$ (GPU) $=48$. 

\subsection{Baseline Models}
In our study, we examine three existing models: DiffSound by \citet{yang2022diffsound}, AudioGen by \citet{Kreuk2022AudioGenTG}, and AudioLDM by \citet{Liu2023AudioLDMTG}. AudioGen and DiffSound use text embeddings for conditional generative training, while AudioLDM employs audio embeddings to avoid potential noise from weak textual descriptions in the paired text-audio data. AudioLDM uses audio embeddings from CLAP and asserts that they are effective in capturing cross-modal information. The models were pre-trained on large datasets, including AudioSet, and fine-tuned on the AudioCaps dataset, before evaluation, for enhanced performance. Thus, comparing them to our model \model{} would not be entirely fair.

Despite being trained on a much smaller dataset, our model \model{} outperformed the baselines that were trained on significantly larger datasets. We may largely attribute this to the use of LLM \textsc{Flan-T5}. Therefore, our model \model{} sets itself apart from the three existing models, making it an exciting addition to the current research in this area.

It is important to note that the \texttt{AudioLDM-L-Full-FT} checkpoint from \citet{Liu2023AudioLDMTG} was not available for our study. Therefore, we used the \texttt{AudioLDM-M-Full-FT} checkpoint, which was released by the authors and has $416$M parameters. This checkpoint was fine-tuned on both the AudioCaps and MusicCaps datasets. We performed a subjective evaluation using this checkpoint in our study. We attempted to fine-tune the \texttt{AudioLDM-L-Full} checkpoint on the AudioCaps dataset. However, we were unable to reproduce the results reported in \citet{Liu2023AudioLDMTG} due to a lack of information on the hyperparameters used.

Our model can be compared directly to \texttt{AudioLDM-L} since it has almost the same number of parameters and was trained solely on the AudioCaps dataset. However, it is worth noting that \citet{Liu2023AudioLDMTG} did not release this checkpoint, which made it impossible for us to conduct a subjective evaluation of its generated samples.
\subsection{Evaluation Metrics}
\paragraph{Objective Evaluation.}
 In this work, we used two commonly used objective metrics: Frechet Audio Distance (FAD) and KL divergence. FAD~\cite{kilgour2019frechet} is a perceptual metric that is adapted from Fechet Inception Distance (FID) for the audio domain. Unlike reference-based metrics, it measures the distance between the generated audio distribution and the real audio distribution without using any reference audio samples. On the other hand, KL divergence~\cite{yang2022diffsound,Kreuk2022AudioGenTG} is a reference-dependent metric that computes the divergence between the distributions of the original and generated audio samples based on the labels generated by a pre-trained classifier. While FAD is more related to human perception, KL divergence captures the similarities between the original and generated audio signals based on broad concepts present in them. In addition to FAD, we also used Frechet Distance (FD)~\cite{Liu2023AudioLDMTG} as an objective metric. FD is similar to FAD, but it replaces the VGGish classifier with PANN. The use of different classifiers in FAD and FD allows us to evaluate the performance of the generated audio using different feature representations.
\paragraph{Subjective Evaluation.}

Following \citet{Liu2023AudioLDMTG} and \citet{Kreuk2022AudioGenTG}, we ask six human evaluators to assess two aspects –– overall audio quality (OVL) and relevance to the input text (REL) -- of 30 randomly selected baseline- and \model{}-generated audio samples on a scale from 1 to 100. The evaluators were proficient in the English language and instructed well to make a fair assessment.

\subsection{Results and Analysis}
\paragraph{Main Results.}
We report our main comparative study in \Cref{tab:AudioCapsResults}. We comapre our proposed method \model{} with DiffSound~\cite{yang2022diffsound}, AudioGen~\cite{Kreuk2022AudioGenTG} and various configurations of AudioLDM~\cite{Liu2023AudioLDMTG}. AudioLDM obtained best results with 200 sampling steps from the LDM during inference. For a fair comparison, we also use 200 inference steps in \model{} and in our additional AudioLDM experiments. We used a classifier-free guidance scale of 3 for \model{}. AudioLDM used a guidance scale among \{2, 2.5, 3\} in their various experiments.

\model{} achieves new state-of-the-art results for objective metrics when trained only on the AudioCaps dataset, with scores of 24.52 FD, 1.37 KL, and 1.59 FAD. This is significantly better than the most direct baseline \texttt{AudioLDM-L}, which also used only the AudioCaps dataset for LDM training. We attribute this to the use of \textsc{Flan-T5} as text encoder in \model{}. We also note that \model{} matches or beats the performance of \texttt{AudioLDM-*-FT} models, which used significantly ($\sim$ 63 times) larger datasets for LDM training. The \texttt{AudioLDM-*-FT} models used two phases of LDM training -- first on the collection of the four datasets, and then only on AudioCaps. \model{} is thus far more sample efficient as compared to the \texttt{AudioLDM-*-FT} model family.

\model{} also shows very promising results for subjective evaluation, with an overall audio quality score of 85.94 and a relevance score of 80.36, indicating its significantly better audio generation ability compared to AudioLDM and other baseline text-to-audio generation approaches. 

\begin{table*}[ht!]
\centering
\small
\caption{The comparison between \model{} and baseline TTA models. \emph{FT} indicates the model is fine-tuned on the Audiocaps (AC) dataset. The AS and AC stand for AudioSet and AudiocCaps datasets respectively. We borrowed all the results from \cite{Liu2023AudioLDMTG} except for AudioLDM-L-Full which was evaluated using the model released by the authors on Huggingface. Despite the LDM being trained on a much smaller dataset, \model{} outperforms AudioLDM and other baseline TTA models as per both objective and subjective metrics. $^\ddag$ indicates the results are obtained using the checkpoints released by \citet{Liu2023AudioLDMTG}.}
\resizebox{\textwidth}{!}{
\begin{tabular}{cccc|ccc|cc}
\toprule
\multirow{2}{*}{\textbf{Model}} & \multirow{2}{*}{\textbf{Datasets}} & \multirow{2}{*}{\textbf{Text}} & \multirow{2}{*}{\textbf{\#Params}} & \multicolumn{3}{c}{\textbf{Objective Metrics}} & \multicolumn{2}{|c}{\textbf{Subjective Metrics}} \\ 
& & & & FD~$\downarrow$ & KL~$\downarrow$ & FAD~$\downarrow$ & OVL~$\uparrow$ & REL~$\uparrow$ \\
\midrule
Ground truth & $-$ & $-$ & $-$ & $-$ & $-$ & $-$ & $91.61$ & $86.78$ \\
\midrule
DiffSound~\cite{yang2022diffsound}     & AS+AC & \greencheck          & $400$M   & $47.68$ & $2.52$ & $7.75$ & $-$ & $-$ \\
AudioGen~\cite{Kreuk2022AudioGenTG}      & AS+AC+8 others & \greencheck &$285$M  &  $-$    & $2.09$  & $3.13$ & $-$ & $-$ \\
AudioLDM-S & AC      & 
\redcross       & $181$M   & $29.48$  & $1.97$ & $2.43$ & $-$ & $-$ \\
AudioLDM-L & AC       & 
\redcross      & $739$M  & $27.12$  & $1.86$ & $2.08$ 
 & $-$ & $-$\\
 \\
AudioLDM-M-Full-FT$^\ddag$ & AS+AC+2 others & 
\redcross & $416$M  & ${26.12}$  & $\mathbf{1.26}$ & ${2.57}$ & ${79.85}$ & ${76.84}$ \\
 AudioLDM-L-Full$^\ddag$ & AS+AC+2 others & 
\redcross & $739$M  & ${32.46}$   & ${1.76}$ & ${4.18}$ & ${78.63}$ & ${62.69}$ \\
AudioLDM-L-Full-FT & AS+AC+2 others & 
\redcross & $739$M  & $\mathbf{23.31}$  & ${1.59}$ & ${1.96}$ & ${-}$ & ${-}$ \\
\midrule
\model{} & AC & 
\greencheck & $866$M  & $24.52$   & 1.37 & $\mathbf{1.59}$ & $\mathbf{85.94}$ & $\mathbf{80.36}$ \\
\bottomrule
\end{tabular}
}
\label{tab:AudioCapsResults}
\end{table*}

\begin{table*}[ht!]
\centering
\caption{The comparison between \model{} and baseline TTA models when trained on the corpus of large datasets. \model{}-Full-FT was first pre-trained on a corpus comprising samples from AudioSet, AudioCaps, Freesound, and BBC datasets followed by fine-tuning on AudioCaps.}
\resizebox{\textwidth}{!}{
\begin{tabular}{ccccc|cc}
\toprule
\multirow{2}{*}{\textbf{Model}} & \multirow{2}{*}{\textbf{Datasets}} & \multirow{2}{*}{\textbf{Dataset Size}} & \multirow{2}{*}{\textbf{Text}} & \multirow{2}{*}{\textbf{\#Params}} & \multicolumn{2}{c}{\textbf{Objective Metrics}} \\ 
&  && & & FD~$\downarrow$ & KL~$\downarrow$ \\
\midrule
AudioLDM-M-Full-FT$^\ddag$ & AS+AC+2 others & 3.3M & 
\redcross & $416$M  & ${26.12}$  & $1.26$ \\
 AudioLDM-L-Full$^\ddag$ & AS+AC+2 others & 3.3M & 
\redcross & $739$M  & ${32.46}$   & ${1.76}$  \\
AudioLDM-L-Full-FT & AS+AC+2 others & 3.3M & 
\redcross & $739$M  & $23.31$  & ${1.59}$ \\
\midrule
\model{}\textsc{-Full-FT} & AS+AC+7 others & 1.2M & 
\greencheck & $866$M  & $\mathbf{18.93}$   & $\mathbf{1.12}$ \\
\bottomrule
\end{tabular}
}
\label{tab:AudioCapsResults-ft}
\end{table*}

\paragraph{Training on Larger Datasets.}
In this experiment, we followed a two-step process to enhance the performance of \model{}. First, we conducted pre-training using a diverse corpus consisting of textual prompts and audio samples sourced from WavCaps~\cite{mei2023wavcaps}, AudioCaps, ESC~\cite{piczak2015dataset}, UrbanSound~\cite{salamon2014dataset}, MusicCaps~\cite{agostinelli2023musiclm}, GTZAN~\cite{tzanetakis2002musical}, and Musical Instruments\footnote{https://www.kaggle.com/datasets/soumendraprasad/musical-instruments-sound-dataset} dataset. The dataset statistics are reported \cref{tab:dataset-statistics}. All audio clips of longer than 10 seconds were segmented into partitions of successive 10 seconds or shorter. We also resampled all audio clips to 16KHz.
The WavCaps dataset consists of ChatGPT-generated captions for the FreeSound\footnote{https://freesound.org/}, BBC Sound Effects\footnote{https://sound-effects.bbcrewind.co.uk} (SFX), and the AudioSet strongly labeled subset. The Urban Sound and ESC50 datasets contain various environmental sounds. The Musical Instruments dataset contains sounds of guitar, drum, violin, and piano instruments. The GTZAN dataset contains sounds of different musical genres -- classical, jazz, etc. These four datasets -- Urban Sound, ESC50, Musical Instruments, GTZAN are audio classification datasets. We use the classification label e.g. \textit{piano} and a more natural prompt \textit{sound of piano} to create two different training instances for each audio sample for these datasets. 

The initial pre-training stage aimed to capture a broad understanding of audio and text interactions. Next, we fine-tuned the pre-trained model specifically on the AudioCaps dataset. The obtained results, as presented in \cref{tab:AudioCapsResults-ft}, demonstrate a remarkable performance improvement achieved by \model{}\textsc{-Full-FT} compared to similar models in the AudioLDM family. These comparable models underwent identical pre-training and fine-tuning approaches, highlighting the effectiveness of our methodology in enhancing the model's overall performance. We conducted pre-training on \model{} for a duration of $200,000$ steps using four A6000 GPUs. To optimize the training process, we set the batch size per GPU to $2$ and employed $8$ gradient accumulation steps, which effectively increased the batch size to $64$. We fine-tuned the model on AudioCaps for $57K$ steps. To help open-source research in TTA, we released this dataset publicly~\footnote{\url{https://huggingface.co/datasets/declare-lab/TangoPromptBank}}.
\begin{table}[htbp]
\centering
\caption{Statistics of the datasets used in training \model{}\textsc{-Full-FT}.}
\label{tab:dataset-statistics}
\resizebox{\textwidth}{!}{
\begin{tabular}{lccccccccc|c}
\toprule
\textbf{Model} & \textbf{AudioSet} & \textbf{AudioCaps} & \textbf{Freesound} & \textbf{BBC} & \begin{tabular}{c}
     \textbf{Urban}   \\
     \textbf{Sound}
\end{tabular} & \begin{tabular}{c}
     \textbf{Musical}   \\
     \textbf{Instrument}
\end{tabular} & \textbf{MusicCaps} & \begin{tabular}{c}
     \textbf{Gtzan}   \\
     \textbf{Music Genre}
\end{tabular} & \textbf{ESC50}  & \textbf{Total} \\
\midrule
\model{}        & $108K$        & $45K$        & $680K$         & $374K$ & $17K$ & $12K$ & $10K$ & $6K$ & $4K$ & $1.2M$   \\
AudioLDM        & $2.1M$        & $49K$         & $680K$         & $374K$ & - & - & - & - & -& $3.3M$   \\
\bottomrule
\end{tabular}
}
\end{table}

\paragraph{Effect of Different Data Augmentation Strategies.}
Table \ref{tab:augmentation} presents a comparison between random and relative pressure-based data augmentation strategies. Notably, the relative pressure-based augmentation strategy yields the most promising results. When evaluating \model{} against AudioLDM-L, both utilizing random data augmentation strategies, \model{} outperforms AudioLDM-L in two out of three objective metrics. This notable improvement can be attributed to the integration of a powerful large language model (\textsc{flan-t5}) as a textual prompt encoder within \model{}.

\begin{table}[ht!]
\centering
\caption{Effect on the objective evaluation metrics with random vs. relative pressure-guided augmentation. Scores were computed for a guidance scale of 3 and 200 inference steps.}
\begin{tabular}{c|cccc}
\toprule
Model & Augmentation & FD~$\downarrow$ & KL~$\downarrow$ & FAD~$\downarrow$ \\
\midrule
\multirow{2}{*}{\model{}} & Random & $25.84$ & $1.38$ & $2.72$ \\
& Relative Pressure & $\mathbf{24.52}$ & $\mathbf{1.37}$ & $\mathbf{1.59}$ \\
\midrule
AudioLDM-L & Random & $27.12$  & $1.86$ & $2.08$ \\
\bottomrule
\end{tabular}
\label{tab:augmentation}
\end{table}

\paragraph{Effect of Inference Steps and Classifier-Free Guidance.}
The number of inference steps and the classifier-free guidance scale are of crucial importance for sampling from latent diffusion models~\cite{Song2020DenoisingDI,Ho2022ClassifierFreeDG}. We report the effect of varying number of steps and varying guidance scale for audio generation in AudioCaps in \Cref{tab:Steps&Guidance}. We found that a guidance scale of 3 provides the best results for \model{}. In the left part of \Cref{tab:Steps&Guidance}, we fix the guidance scale of 3 and vary the number of steps from 10 to 200. The generated audio quality and resultant objective metrics consistently become better with more steps. ~\citet{Liu2023AudioLDMTG} reported that the performance for AudioLDM plateaus at around 100 steps, with 200 steps providing only marginally better performance. However, we notice a substantial improvement in performance when going from 100 to 200 inference steps for \model{}, suggesting that there could be further gain in performance with more inference steps. 

\begin{table*}[ht!]
\centering
\small
\caption{Effect on the objective evaluation metrics with a varying number of inference steps and classifier-free guidance.}
\resizebox{\textwidth}{!}{
\begin{tabular}{c|ccccc|ccccc}
\toprule
\multirow{2}{*}{\textbf{Model}} & \multicolumn{5}{c|}{\textbf{Varying Steps}} & \multicolumn{5}{c}{\textbf{Varying Guidance}} \\
& Guidance & Steps & FD~$\downarrow$ & KL~$\downarrow$ & FAD~$\downarrow$ & Steps & Guidance & FD~$\downarrow$ & KL~$\downarrow$ & FAD~$\downarrow$ \\
\midrule

\multirow{5}{*}{\model{}} & \multirow{5}{*}{$3$} & $10$ & $45.12$ & $1.66$ & $11.38$ & \multirow{5}{*}{$100$} & - & 35.76 & 2.02 & 6.22 \\
& & $20$ & $31.38$ & $1.39$ & $4.52$ & & $2.5$ & $26.32$ & $1.39$ & $1.97$ \\
& & $50$ & $25.33$ & $\mathbf{1.27}$ & $2.13$ & & 3 & $26.13$ & $1.37$ & $\mathbf{1.87}$ \\
& & $100$ & $26.13$ & $1.37$ & $1.87$ & & $5$ & $\mathbf{24.28}$ & $\mathbf{1.28}$ & $2.32$ \\
& & $200$ & $\mathbf{24.52}$ & $1.37$ & $\mathbf{1.59}$ & & $10$ & $26.10$ & $1.31$ & $3.30$ \\
\bottomrule
\end{tabular}
}
\label{tab:Steps&Guidance}
\end{table*}

We report the effect of varying guidance scale with a fixed 100 steps in the right half of \Cref{tab:Steps&Guidance}. The first row uses a guidance scale of 1, thus effectively not applying classifier-free guidance at all during inference. Not surprisingly, the performance of this configuration is poor, lagging far behind the classifier-free guided models across all the objective measures. We obtain almost similar results with a guidance scale of 2.5 and better FD and KL with a guidance scale of 5. We obtain the best FAD metric at a guidance scale of 3 and the metric becomes poorer with larger guidance.

\paragraph{Temporal Sequence Modelling.}
We analyze how \model{} and AudioLDM models perform audio generation when the text prompt contains multiple sequential events. Consider the following examples: \textit{\small{\underline{A toy train running as a young boy talks} followed by \underline{plastic clanking} then \underline{a child laughing}}} contains three separate sequential events, whereas \textit{\small{Rolling thunder with lightning strikes}} contains only one. We segregate the AudioCaps test set using the presence of temporal identifiers -- \textit{while, before, after, then, followed} -- into two subsets, one with multiple events and the other with single event. We show the objective evaluation results for audio generation on these subsets in \Cref{tab:TemporalAnalysis}. \model{} achieves the best FD and FAD scores for both multiple events and single event instances. The best KL divergence score is achieved by the \texttt{AudioLDM-M-Full-FT} model. We conjecture that the larger corpus from the four training datasets in AudioLDM could be more helpful in improving the reference-based KL metric, unlike the reference-free FD and FAD metrics. 
\begin{table*}[ht!]
\centering
\small
\caption{Objective evaluation results for audio generation in the presence of multiple events or a single event in the text prompt in the AudioCaps test set. The multiple events and single event subsets collectively constitute the entire AudioCaps test set. It should be noted that FD and FAD are corpus-level non-linear metrics, and hence the FD and FAD scores reported in \cref{tab:AudioCapsResults} are not average of the subset scores reported in this table.}
\resizebox{0.95\textwidth}{!}{
\begin{tabular}{cc|ccc|ccc}
\toprule
\multirow{2}{*}{\textbf{Model}} & \multirow{2}{*}{\textbf{Datasets}} & \multicolumn{3}{c|}{\textbf{Multiple Events}} & \multicolumn{3}{c}{\textbf{Single Event}} \\

& & FD~$\downarrow$ & KL~$\downarrow$ & FAD~$\downarrow$ & FD~$\downarrow$ & KL~$\downarrow$ & FAD~$\downarrow$ \\
\midrule
AudioLDM-L-Full & \multirow{2}{*}{AS+AC+2 others} & $43.65$ & $1.90$ & $3.77$ & $35.39$ & $1.66$ & $5.24$ \\
AudioLDM-M-Full-FT & & $34.57$ & $\mathbf{1.32}$ & $2.45$ & $29.40$ & $\mathbf{1.21}$ & $3.27$ \\
\midrule
\model{} & AC & $\mathbf{33.36}$ & $1.45$ & $\mathbf{1.75}$ & $\mathbf{28.59}$ & $1.30$ & $\mathbf{2.04}$\\
\bottomrule
\end{tabular}
}
\label{tab:TemporalAnalysis}
\end{table*}

\paragraph{Performance against Number of Labels.}
Recall that the AudioCaps dataset was curated from the annotations of the audio classification task in the AudioSet dataset. The text prompts in AudioCaps can thus be paired with the discrete class labels of AudioSet. The AudioSet dataset contains a total of 632 audio event classes. For instance, \textit{\small{A woman and a baby are having a conversation}} and its corresponding audio clip has the following three labels: \textit{\small{Speech, Child speech kid speaking, Inside small room}}. We group instances having one label, two labels, and multiple (two or more) labels in AudioCaps and evaluate the generated audios across the objective metrics. We report the result of the experiment in \Cref{tab:LabelAnalysis}. \model{} outperforms AudioLDM models across all the objective metrics for audio generation from texts with one label or two labels. For texts with multiple labels, AudioLDM achieves a better KL divergence score and \model{} achieves better FD and FAD scores. Interestingly, all the models achieve consistently better FD and KL scores with progressively more labels, suggesting that such textual prompts are more effectively processed by the diffusion models.
\begin{table*}[ht!]
\centering
\small
\caption{Performance of audio generation in AudioCaps for texts containing one, two, or multiple (two or more) labels. Each text in AudioCaps has its corresponding multi-category labels from AudioSet. We use these labels to segregate the AudioCaps dataset into three subsets.}
\resizebox{\textwidth}{!}{
\begin{tabular}{cc|ccc|ccc|ccc}
\toprule
\multirow{2}{*}{\textbf{Model}} & \multirow{2}{*}{\textbf{Datasets}} & \multicolumn{3}{c|}{\textbf{One Label}} & \multicolumn{3}{c|}{\textbf{Two Labels}} & \multicolumn{3}{c}{\textbf{Multiple Labels}} \\

& & FD~$\downarrow$ & KL~$\downarrow$ & FAD~$\downarrow$ & FD~$\downarrow$ & KL~$\downarrow$ & FAD~$\downarrow$ & FD~$\downarrow$ & KL~$\downarrow$ & FAD~$\downarrow$ \\
\midrule
AudioLDM-L-Full & \multirow{2}{*}{AS+AC+2 others} & $48.11$ & $2.07$ & $4.71$ & $44.93$ & $1.90$ & $4.09$ & 34.94 & 1.68 & 4.59 \\
AudioLDM-M-Full-FT & & 46.44 & $1.85$ & $3.77$ & $39.01$ & $1.29$ & $3.52$ & $26.74$ & $\mathbf{1.10}$ & $2.62$ \\
\midrule
\model{} & AC & $\mathbf{40.81}$ & $\mathbf{1.84}$ & $\mathbf{1.79}$ & $\mathbf{35.09}$ & $\mathbf{1.56}$ & $\mathbf{2.53}$ & $\mathbf{26.05}$ & $1.24$ & $\mathbf{1.96}$ \\
\bottomrule
\end{tabular}
 }
\label{tab:LabelAnalysis}
\end{table*}

\begin{figure}[ht]
    \centering
    \includegraphics[width=0.8\textwidth]{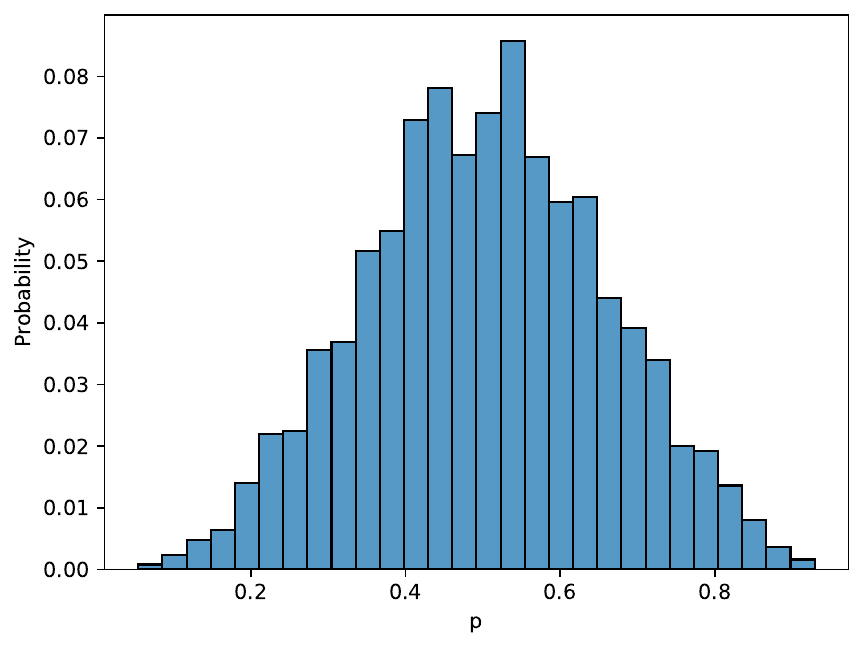}
    \caption{Distribution of relative pressure level (see \Cref{eq:gain}) across the augmented samples.}
    \label{fig:relative_gain_dist}
\end{figure}

\paragraph{Effect of Augmentation and Distribution of Relative Pressure-Level ($p$) for Augmentation}
We described our augmentation strategy earlier in \Cref{sec:augmentation}. The distribution of the relative pressure level $p$ in \Cref{eq:gain} across the training samples is shown in \Cref{fig:relative_gain_dist} that implies that the relative pressure levels are roughly normally distributed and many samples have low levels of relative pressure, which might be poorly represented in a random mixing. In contrast, our approach allows for much equitable mixing.

\begin{table*}[ht!]
\small
\centering
\caption{Performance of  \texttt{AudioLDM-M-Full FT} and \model{} for the most frequently occurring categories in AudioCaps dataset. CEB indicates the Channel, environment, and background sounds category.}
\resizebox{\linewidth}{!}{
\begin{tabular}{ccccc|cc|cc|cc}
\toprule
\multirow{2}{*}{Human} & \multirow{2}{*}{Animal} & \multirow{2}{*}{Natural} & \multirow{2}{*}{Things} & \multirow{2}{*}{CEB} & \multicolumn{2}{c|}{FD~$\downarrow$} & \multicolumn{2}{c|}{KL~$\downarrow$} & \multicolumn{2}{c}{FAD~$\downarrow$} \\
& & & & & AudioLDM & \model{} & AudioLDM & \model{} & AudioLDM & \model{}\\
\midrule
\greencheck & \redcross & \redcross & \redcross & \redcross & $38.15$ & $\mathbf{34.06}$ & $1.01$ & $\mathbf{0.99}$ & $2.81$ & $\mathbf{2.13}$ \\
\redcross & \greencheck & \redcross & \redcross & \redcross & $78.62$ & $\mathbf{77.78}$ & $\mathbf{1.82}$ & $1.92$ & $\mathbf{4.28}$ & $4.62$ \\
\greencheck & \greencheck & \redcross & \redcross & \redcross & $\mathbf{61.91}$ & $70.32$ & $\mathbf{0.89}$ & $1.29$ & $6.32$ & $\mathbf{5.19}$ \\
\redcross & \redcross & \greencheck & \redcross & \redcross & $\mathbf{51.61}$ & $57.75$ & $\mathbf{1.89}$ & $1.96$ & $6.75$ & $\mathbf{5.15}$ \\
\redcross & \redcross & \redcross & \greencheck & \redcross & $35.60$ & $\mathbf{33.13}$ & $\mathbf{1.35}$ & $1.43$ & $5.42$ & $\mathbf{3.40}$ \\
\redcross & \redcross & \greencheck & \greencheck & \redcross & $55.06$ & $\mathbf{42.00}$ & $1.46$ & $\mathbf{1.12}$ & $6.57$ & $\mathbf{3.89}$ \\
\greencheck & \redcross & \redcross & \greencheck & \redcross & $\mathbf{37.57}$ & $39.22$ & $\mathbf{1.11}$ & $1.34$ & $3.26$ & $\mathbf{3.18}$ \\
\redcross & \redcross & \redcross & \greencheck & \greencheck & $54.25$ & $\mathbf{52.77}$ & $1.43$ & $\mathbf{1.33}$ & $11.49$ & $\mathbf{9.26}$ \\

\bottomrule
\end{tabular}
}
\label{tab:categories}

\end{table*}
\paragraph{Categorical Modelling.}
The class labels in AudioSet can be arranged hierarchically to obtain the following top-level categories: i) Human sounds, ii) Animal sounds, iii) Natural sounds, iv) Sounds of Things, v) Channel, environment, background sounds, vi) Source-ambiguous sounds, and vii) Music. We map the class labels in AudioCaps to the seven main categories listed above. The Music category is very rare in AudioCaps and the rest either appear on their own or in various combinations with others. We select the most frequently occurring category combinations and analyze the performance of various models within the constituting AudioCaps instances in \Cref{tab:categories}. The performance of the two models is pretty balanced across the FD and KL metrics, with \model{} being better in some, and AudioLDM in others. However, \model{} achieves better FAD scores in all but one group, with large improvements in (human, animal), (natural), (things), and (natural, things) categories.

\section{Related Works}
\paragraph{Diffusion Models.}
Recent years have seen a surge in diffusion models as a leading approach for generating high-quality speech \cite{chen2020wavegrad,kong2020diffwave,popov2021grad,popov2021diffusion,jeong2021diff,huang2022fastdiff}. These models utilize a fixed number of Markov chain steps to transform white noise signals into structured waveforms. Among them, FastDiff has achieved remarkable results in high-quality speech synthesis \cite{huang2022fastdiff}. By leveraging a stack of time-aware diffusion processes, FastDiff can generate speech samples of exceptional quality at an impressive speed, 58 times faster than real-time on a V100 GPU, making it practical for speech synthesis deployment. It surpasses other existing methods in end-to-end text-to-speech synthesis. Another noteworthy probabilistic model for audio synthesis is DiffWave \cite{kong2020diffwave}, which is non-autoregressive and generates high-fidelity audio for various waveform generation tasks, including neural vocoding conditioned on mel spectrogram, class-conditional generation, and unconditional generation. DiffWave delivers speech quality that is on par with the powerful WaveNet vocoder \cite{oord2016wavenet} while synthesizing audio much faster.
Diffusion models have emerged as a promising approach for speech processing, particularly in speech enhancement \cite{9689602,serra2022universal,qiu2022srtnet,9746901}. Recent advancements in diffusion probabilistic models have led to the development of a new speech enhancement algorithm that incorporates the characteristics of noisy speech signals into the forward and reverse diffusion processes \cite{lu2022conditional}. This new algorithm is a generalized form of the probabilistic diffusion model, known as the conditional diffusion probabilistic model. During its reverse process, it can adapt to non-Gaussian real noises in the estimated speech signal, making it highly effective in improving speech quality.
In addition, \citet{qiu2022srtnet} propose SRTNet, a novel method for speech enhancement that incorporates the diffusion model as a module for stochastic refinement. The proposed method comprises a joint network of deterministic and stochastic modules, forming the “enhance-and-refine” paradigm. The paper also includes a theoretical demonstration of the proposed method’s feasibility and presents experimental results to support its effectiveness, highlighting its potential in improving speech quality. 

\paragraph{Text-to-Audio Generation.}
The field of text-to-audio generation has received limited attention until recently~\cite{Kreuk2022AudioGenTG,yang2022diffsound}. In \citet{yang2022diffsound}, a text encoder is used to obtain text features, which are then processed by a non-autoregressive decoder to generate spectrogram tokens. These tokens are fed to a vector quantized VAE (VQ-VAE) to generate mel spectrograms that are used by a vocoder to generate audio. The non-autoregressive decoder is a probabilistic diffusion model. In addition, \citet{yang2022diffsound} introduced a novel data augmentation technique called the mask-based text generation strategy (MBTG), which masks out portions of input text that do not represent any event, such as those indicating temporality. The aim of MBTG is to learn augmented text descriptions from audio during training. Although this approach seems promising, its fundamental limitation is the lack of diversity in the generated data, as it fails to mix different audio samples. Later, \citet{Kreuk2022AudioGenTG} proposed a correction to this method, mixing audio signals according to random signal-to-noise ratios and concatenating the corresponding textual descriptions. This approach allows for the generation of new (text, audio) pairs and mitigates the limitations of \citet{yang2022diffsound}. Unlike \citet{yang2022diffsound}, the architecture proposed in \citet{Kreuk2022AudioGenTG} uses a transformer encoder and decoder network to autoregressively generate audio tokens from text input.

Recently, \citet{Liu2023AudioLDMTG} proposed AudioLDM, which translates the Latent Diffusion Model of text-to-visual to text-to-audio generation. They pre-trained VAE-based encoder-decoder networks to learn a compressed latent representation of audio, which was then used to guide a diffusion model to generate audio tokens from text input. They found that using audio embeddings instead of text embeddings during the backward diffusion process improved conditional audio generation. During inference time, they used text embeddings for text-to-audio generation. Audio and text embeddings were obtained using pre-trained CLAP, which is the audio counterpart of CLIP embeddings used in the original LDM model.

\section{Limitations}
\model{} is not always able to finely control its generations over textual control prompts as it is trained only on the small AudioCaps dataset. For example, the generations from \model{} for prompts \textit{Chopping tomatoes on a wooden table} and \textit{Chopping potatoes on a metal table} are very similar. \textit{Chopping vegetables on a table} also produces similar audio samples. Training text-to-audio generation models on larger datasets is thus required for the model to learn the composition of textual concepts and varied text-audio mappings. In the future, we plan to improve \model{} by training it on larger datasets and enhancing its compositional and controllable generation ability.

\section{Conclusion}
In this work, we investigate the effectiveness of the instruction-tuned model, \textsc{Flan-T5}, for text-to-audio generation. Specifically, we use the textual embeddings produced by \textsc{Flan-T5} in the latent diffusion model to generate mel-spectrogram tokens. These tokens are then fed to a pre-trained variational auto-encoder (VAE) to generate mel-spectrograms, which are later used by a pre-trained vocoder to generate audio. Our model achieved superior performance under both objective and subjective evaluations compared to the state-of-the-art text-to-audio model, AudioLDM, despite using only 63 times less training data. We primarily attribute this performance improvement to the representational power of \textsc{Flan-T5}, which is due to its instruction-based tuning in the pre-training stage. In the future, we plan to investigate the effectiveness of \textsc{Flan-T5} in other audio tasks, such as, audio super-resolution and inpainting.

\section*{Acknowledgements}
We are grateful to \textsc{Oracle for Research} and \textsc{Huggingface} for their generous support to the project \model{}. 
\bibliographystyle{plainnat}
\bibliography{refs}

\begin{thebibliography}{43}
\providecommand{\natexlab}[1]{#1}
\providecommand{\url}[1]{\texttt{#1}}
\expandafter\ifx\csname urlstyle\endcsname\relax
  \providecommand{\doi}[1]{doi: #1}\else
  \providecommand{\doi}{doi: \begingroup \urlstyle{rm}\Url}\fi

\bibitem[Agostinelli et~al.(2023)Agostinelli, Denk, Borsos, Engel, Verzetti,
  Caillon, Huang, Jansen, Roberts, Tagliasacchi,
  et~al.]{agostinelli2023musiclm}
Andrea Agostinelli, Timo~I Denk, Zal{\'a}n Borsos, Jesse Engel, Mauro Verzetti,
  Antoine Caillon, Qingqing Huang, Aren Jansen, Adam Roberts, Marco
  Tagliasacchi, et~al.
\newblock Musiclm: Generating music from text.
\newblock \emph{arXiv preprint arXiv:2301.11325}, 2023.

\bibitem[Chen et~al.(2020)Chen, Zhang, Zen, Weiss, Norouzi, and
  Chan]{chen2020wavegrad}
Nanxin Chen, Yu~Zhang, Heiga Zen, Ron~J Weiss, Mohammad Norouzi, and William
  Chan.
\newblock Wavegrad: Estimating gradients for waveform generation.
\newblock \emph{arXiv preprint arXiv:2009.00713}, 2020.

\bibitem[Chung et~al.(2022)Chung, Hou, Longpre, Zoph, Tay, Fedus, Li, Wang,
  Dehghani, Brahma, Webson, Gu, Dai, Suzgun, Chen, Chowdhery, Narang, Mishra,
  Yu, Zhao, Huang, Dai, Yu, Petrov, Chi, Dean, Devlin, Roberts, Zhou, Le, and
  Wei]{https://doi.org/10.48550/arxiv.2210.11416}
Hyung~Won Chung, Le~Hou, Shayne Longpre, Barret Zoph, Yi~Tay, William Fedus,
  Eric Li, Xuezhi Wang, Mostafa Dehghani, Siddhartha Brahma, Albert Webson,
  Shixiang~Shane Gu, Zhuyun Dai, Mirac Suzgun, Xinyun Chen, Aakanksha
  Chowdhery, Sharan Narang, Gaurav Mishra, Adams Yu, Vincent Zhao, Yanping
  Huang, Andrew Dai, Hongkun Yu, Slav Petrov, Ed~H. Chi, Jeff Dean, Jacob
  Devlin, Adam Roberts, Denny Zhou, Quoc~V. Le, and Jason Wei.
\newblock Scaling instruction-finetuned language models, 2022.
\newblock URL \url{https://arxiv.org/abs/2210.11416}.

\bibitem[Dai et~al.(2022)Dai, Sun, Dong, Hao, Sui, and Wei]{Dai2022WhyCG}
Damai Dai, Yutao Sun, Li~Dong, Yaru Hao, Zhifang Sui, and Furu Wei.
\newblock Why can gpt learn in-context? language models secretly perform
  gradient descent as meta-optimizers.
\newblock \emph{ArXiv}, abs/2212.10559, 2022.

\bibitem[Gemmeke et~al.(2017)Gemmeke, Ellis, Freedman, Jansen, Lawrence, Moore,
  Plakal, and Ritter]{gemmeke2017audio}
Jort~F Gemmeke, Daniel~PW Ellis, Dylan Freedman, Aren Jansen, Wade Lawrence,
  R~Channing Moore, Manoj Plakal, and Marvin Ritter.
\newblock Audio set: An ontology and human-labeled dataset for audio events.
\newblock In \emph{2017 IEEE international conference on acoustics, speech and
  signal processing (ICASSP)}, pages 776--780. IEEE, 2017.

\bibitem[Hang et~al.(2023)Hang, Gu, Li, Bao, Chen, Hu, Geng, and
  Guo]{hang2023efficient}
Tiankai Hang, Shuyang Gu, Chen Li, Jianmin Bao, Dong Chen, Han Hu, Xin Geng,
  and Baining Guo.
\newblock Efficient diffusion training via min-snr weighting strategy, 2023.

\bibitem[Ho and Salimans(2021)]{Ho2022ClassifierFreeDG}
Jonathan Ho and Tim Salimans.
\newblock Classifier-free diffusion guidance.
\newblock \emph{NeurIPS 2021 Workshop on Deep Generative Models and Downstream
  Applications}, 2021.

\bibitem[Huang et~al.(2022)Huang, Lam, Wang, Su, Yu, Ren, and
  Zhao]{huang2022fastdiff}
Rongjie Huang, Max~WY Lam, Jun Wang, Dan Su, Dong Yu, Yi~Ren, and Zhou Zhao.
\newblock Fastdiff: A fast conditional diffusion model for high-quality speech
  synthesis.
\newblock \emph{arXiv preprint arXiv:2204.09934}, 2022.

\bibitem[Isola et~al.(2016)Isola, Zhu, Zhou, and
  Efros]{Isola2016ImagetoImageTW}
Phillip Isola, Jun-Yan Zhu, Tinghui Zhou, and Alexei~A. Efros.
\newblock Image-to-image translation with conditional adversarial networks.
\newblock \emph{2017 IEEE Conference on Computer Vision and Pattern Recognition
  (CVPR)}, pages 5967--5976, 2016.

\bibitem[Jeong et~al.(2021)Jeong, Kim, Cheon, Choi, and Kim]{jeong2021diff}
Myeonghun Jeong, Hyeongju Kim, Sung~Jun Cheon, Byoung~Jin Choi, and Nam~Soo
  Kim.
\newblock Diff-tts: A denoising diffusion model for text-to-speech.
\newblock \emph{arXiv preprint arXiv:2104.01409}, 2021.

\bibitem[Kilgour et~al.(2019)Kilgour, Zuluaga, Roblek, and
  Sharifi]{kilgour2019frechet}
Kevin Kilgour, Mauricio Zuluaga, Dominik Roblek, and Matthew Sharifi.
\newblock Fr{\'e}chet audio distance: A reference-free metric for evaluating
  music enhancement algorithms.
\newblock In \emph{INTERSPEECH}, pages 2350--2354, 2019.

\bibitem[Kim et~al.(2019)Kim, Kim, Lee, and Kim]{kim2019audiocaps}
Chris~Dongjoo Kim, Byeongchang Kim, Hyunmin Lee, and Gunhee Kim.
\newblock Audiocaps: Generating captions for audios in the wild.
\newblock In \emph{Proceedings of the 2019 Conference of the North American
  Chapter of the Association for Computational Linguistics: Human Language
  Technologies, Volume 1 (Long and Short Papers)}, pages 119--132, 2019.

\bibitem[Kingma and Welling(2013)]{Kingma2013AutoEncodingVB}
Diederik~P. Kingma and Max Welling.
\newblock Auto-encoding variational bayes.
\newblock \emph{CoRR}, abs/1312.6114, 2013.

\bibitem[Kong et~al.(2020{\natexlab{a}})Kong, Kim, and Bae]{kong2020hifi}
Jungil Kong, Jaehyeon Kim, and Jaekyoung Bae.
\newblock Hifi-gan: Generative adversarial networks for efficient and high
  fidelity speech synthesis.
\newblock \emph{Advances in Neural Information Processing Systems},
  33:\penalty0 17022--17033, 2020{\natexlab{a}}.

\bibitem[Kong et~al.(2021)Kong, Cao, Liu, Choi, and Wang]{Kong2021DecouplingMA}
Qiuqiang Kong, Yin Cao, Haohe Liu, Keunwoo Choi, and Yuxuan Wang.
\newblock Decoupling magnitude and phase estimation with deep resunet for music
  source separation.
\newblock In \emph{International Society for Music Information Retrieval
  Conference}, 2021.

\bibitem[Kong et~al.(2020{\natexlab{b}})Kong, Ping, Huang, Zhao, and
  Catanzaro]{kong2020diffwave}
Zhifeng Kong, Wei Ping, Jiaji Huang, Kexin Zhao, and Bryan Catanzaro.
\newblock Diffwave: A versatile diffusion model for audio synthesis.
\newblock \emph{arXiv preprint arXiv:2009.09761}, 2020{\natexlab{b}}.

\bibitem[Kreuk et~al.(2022)Kreuk, Synnaeve, Polyak, Singer, D'efossez, Copet,
  Parikh, Taigman, and Adi]{Kreuk2022AudioGenTG}
Felix Kreuk, Gabriel Synnaeve, Adam Polyak, Uriel Singer, Alexandre D'efossez,
  Jade Copet, Devi Parikh, Yaniv Taigman, and Yossi Adi.
\newblock Audiogen: Textually guided audio generation.
\newblock \emph{ArXiv}, abs/2209.15352, 2022.

\bibitem[Liu et~al.(2023)Liu, Chen, Yuan, Mei, Liu, Mandic, Wang, and
  Plumbley]{Liu2023AudioLDMTG}
Haohe Liu, Zehua Chen, Yi~Yuan, Xinhao Mei, Xubo Liu, Danilo~P. Mandic, Wenwu
  Wang, and Mark D~. Plumbley.
\newblock Audio{LDM}: Text-to-audio generation with latent diffusion models.
\newblock \emph{ArXiv}, abs/2301.12503, 2023.

\bibitem[Liu et~al.(2019)Liu, Ott, Goyal, Du, Joshi, Chen, Levy, Lewis,
  Zettlemoyer, and Stoyanov]{Liu2019RoBERTaAR}
Yinhan Liu, Myle Ott, Naman Goyal, Jingfei Du, Mandar Joshi, Danqi Chen, Omer
  Levy, Mike Lewis, Luke Zettlemoyer, and Veselin Stoyanov.
\newblock Roberta: A robustly optimized bert pretraining approach.
\newblock \emph{ArXiv}, abs/1907.11692, 2019.

\bibitem[Loshchilov and Hutter(2017)]{loshchilov2017decoupled}
Ilya Loshchilov and Frank Hutter.
\newblock Decoupled weight decay regularization.
\newblock \emph{arXiv preprint arXiv:1711.05101}, 2017.

\bibitem[Lu et~al.(2021)Lu, Tsao, and Watanabe]{9689602}
Yen-Ju Lu, Yu~Tsao, and Shinji Watanabe.
\newblock A study on speech enhancement based on diffusion probabilistic model.
\newblock In \emph{2021 Asia-Pacific Signal and Information Processing
  Association Annual Summit and Conference (APSIPA ASC)}, pages 659--666, 2021.

\bibitem[Lu et~al.(2022{\natexlab{a}})Lu, Wang, Watanabe, Richard, Yu, and
  Tsao]{9746901}
Yen-Ju Lu, Zhong-Qiu Wang, Shinji Watanabe, Alexander Richard, Cheng Yu, and
  Yu~Tsao.
\newblock Conditional diffusion probabilistic model for speech enhancement.
\newblock In \emph{ICASSP 2022 - 2022 IEEE International Conference on
  Acoustics, Speech and Signal Processing (ICASSP)}, pages 7402--7406,
  2022{\natexlab{a}}.
\newblock \doi{10.1109/ICASSP43922.2022.9746901}.

\bibitem[Lu et~al.(2022{\natexlab{b}})Lu, Wang, Watanabe, Richard, Yu, and
  Tsao]{lu2022conditional}
Yen-Ju Lu, Zhong-Qiu Wang, Shinji Watanabe, Alexander Richard, Cheng Yu, and
  Yu~Tsao.
\newblock Conditional diffusion probabilistic model for speech enhancement.
\newblock In \emph{ICASSP 2022-2022 IEEE International Conference on Acoustics,
  Speech and Signal Processing (ICASSP)}, pages 7402--7406. IEEE,
  2022{\natexlab{b}}.

\bibitem[Mei et~al.(2023)Mei, Meng, Liu, Kong, Ko, Zhao, Plumbley, Zou, and
  Wang]{mei2023wavcaps}
Xinhao Mei, Chutong Meng, Haohe Liu, Qiuqiang Kong, Tom Ko, Chengqi Zhao,
  Mark~D Plumbley, Yuexian Zou, and Wenwu Wang.
\newblock Wavcaps: A chatgpt-assisted weakly-labelled audio captioning dataset
  for audio-language multimodal research.
\newblock \emph{arXiv preprint arXiv:2303.17395}, 2023.

\bibitem[Oord et~al.(2016)Oord, Dieleman, Zen, Simonyan, Vinyals, Graves,
  Kalchbrenner, Senior, and Kavukcuoglu]{oord2016wavenet}
Aaron van~den Oord, Sander Dieleman, Heiga Zen, Karen Simonyan, Oriol Vinyals,
  Alex Graves, Nal Kalchbrenner, Andrew Senior, and Koray Kavukcuoglu.
\newblock Wavenet: A generative model for raw audio.
\newblock \emph{arXiv preprint arXiv:1609.03499}, 2016.

\bibitem[Piczak(2015)]{piczak2015dataset}
Karol~J. Piczak.
\newblock {ESC}: {Dataset} for {Environmental Sound Classification}.
\newblock In \emph{Proceedings of the 23rd {Annual ACM Conference} on
  {Multimedia}}, pages 1015--1018. {ACM Press}, 2015.
\newblock ISBN 978-1-4503-3459-4.
\newblock \doi{10.1145/2733373.2806390}.
\newblock URL \url{http://dl.acm.org/citation.cfm?doid=2733373.2806390}.

\bibitem[Popov et~al.(2021{\natexlab{a}})Popov, Vovk, Gogoryan, Sadekova, and
  Kudinov]{popov2021grad}
Vadim Popov, Ivan Vovk, Vladimir Gogoryan, Tasnima Sadekova, and Mikhail
  Kudinov.
\newblock Grad-tts: A diffusion probabilistic model for text-to-speech.
\newblock In \emph{International Conference on Machine Learning}, pages
  8599--8608. PMLR, 2021{\natexlab{a}}.

\bibitem[Popov et~al.(2021{\natexlab{b}})Popov, Vovk, Gogoryan, Sadekova,
  Kudinov, and Wei]{popov2021diffusion}
Vadim Popov, Ivan Vovk, Vladimir Gogoryan, Tasnima Sadekova, Mikhail Kudinov,
  and Jiansheng Wei.
\newblock Diffusion-based voice conversion with fast maximum likelihood
  sampling scheme.
\newblock \emph{arXiv preprint arXiv:2109.13821}, 2021{\natexlab{b}}.

\bibitem[Qiu et~al.(2022)Qiu, Fu, Yu, Yin, Sun, and Huang]{qiu2022srtnet}
Zhibin Qiu, Mengfan Fu, Yinfeng Yu, LiLi Yin, Fuchun Sun, and Hao Huang.
\newblock Srtnet: Time domain speech enhancement via stochastic refinement.
\newblock \emph{arXiv preprint arXiv:2210.16805}, 2022.

\bibitem[Raffel et~al.(2020)Raffel, Shazeer, Roberts, Lee, Narang, Matena,
  Zhou, Li, and Liu]{2020t5}
Colin Raffel, Noam Shazeer, Adam Roberts, Katherine Lee, Sharan Narang, Michael
  Matena, Yanqi Zhou, Wei Li, and Peter~J. Liu.
\newblock Exploring the limits of transfer learning with a unified text-to-text
  transformer.
\newblock \emph{Journal of Machine Learning Research}, 21\penalty0
  (140):\penalty0 1--67, 2020.
\newblock URL \url{http://jmlr.org/papers/v21/20-074.html}.

\bibitem[Ramesh et~al.(2021)Ramesh, Pavlov, Goh, Gray, Voss, Radford, Chen, and
  Sutskever]{Ramesh2021ZeroShotTG}
Aditya Ramesh, Mikhail Pavlov, Gabriel Goh, Scott Gray, Chelsea Voss, Alec
  Radford, Mark Chen, and Ilya Sutskever.
\newblock Zero-shot text-to-image generation.
\newblock \emph{ArXiv}, abs/2102.12092, 2021.

\bibitem[Ramesh et~al.(2022)Ramesh, Dhariwal, Nichol, Chu, and
  Chen]{Ramesh2022HierarchicalTI}
Aditya Ramesh, Prafulla Dhariwal, Alex Nichol, Casey Chu, and Mark Chen.
\newblock Hierarchical text-conditional image generation with clip latents.
\newblock \emph{ArXiv}, abs/2204.06125, 2022.

\bibitem[Rombach et~al.(2022)Rombach, Blattmann, Lorenz, Esser, and
  Ommer]{rombach2022high}
Robin Rombach, Andreas Blattmann, Dominik Lorenz, Patrick Esser, and Bj{\"o}rn
  Ommer.
\newblock High-resolution image synthesis with latent diffusion models.
\newblock In \emph{Proceedings of the IEEE/CVF Conference on Computer Vision
  and Pattern Recognition}, pages 10684--10695, 2022.

\bibitem[Ronneberger et~al.(2015)Ronneberger, Fischer, and
  Brox]{10.1007/978-3-319-24574-4_28}
Olaf Ronneberger, Philipp Fischer, and Thomas Brox.
\newblock U-net: Convolutional networks for biomedical image segmentation.
\newblock In Nassir Navab, Joachim Hornegger, William~M. Wells, and
  Alejandro~F. Frangi, editors, \emph{Medical Image Computing and
  Computer-Assisted Intervention -- MICCAI 2015}, pages 234--241, Cham, 2015.
  Springer International Publishing.
\newblock ISBN 978-3-319-24574-4.

\bibitem[Saharia et~al.(2022)Saharia, Chan, Saxena, Li, Whang, Denton,
  Ghasemipour, Gontijo~Lopes, Karagol~Ayan, Salimans,
  et~al.]{saharia2022photorealistic}
Chitwan Saharia, William Chan, Saurabh Saxena, Lala Li, Jay Whang, Emily~L
  Denton, Kamyar Ghasemipour, Raphael Gontijo~Lopes, Burcu Karagol~Ayan, Tim
  Salimans, et~al.
\newblock Photorealistic text-to-image diffusion models with deep language
  understanding.
\newblock \emph{Advances in Neural Information Processing Systems},
  35:\penalty0 36479--36494, 2022.

\bibitem[Salamon et~al.(2014)Salamon, Jacoby, and Bello]{salamon2014dataset}
Justin Salamon, Christopher Jacoby, and Juan~Pablo Bello.
\newblock A dataset and taxonomy for urban sound research.
\newblock In \emph{Proceedings of the 22nd ACM international conference on
  Multimedia}, pages 1041--1044, 2014.

\bibitem[Serr{\`a} et~al.(2022)Serr{\`a}, Pascual, Pons, Araz, and
  Scaini]{serra2022universal}
Joan Serr{\`a}, Santiago Pascual, Jordi Pons, R~Oguz Araz, and Davide Scaini.
\newblock Universal speech enhancement with score-based diffusion.
\newblock \emph{arXiv preprint arXiv:2206.03065}, 2022.

\bibitem[Song et~al.(2020)Song, Meng, and Ermon]{Song2020DenoisingDI}
Jiaming Song, Chenlin Meng, and Stefano Ermon.
\newblock Denoising diffusion implicit models.
\newblock \emph{ArXiv}, abs/2010.02502, 2020.

\bibitem[Tokozume et~al.(2017)Tokozume, Ushiku, and
  Harada]{DBLP:journals/corr/abs-1711-10282}
Yuji Tokozume, Yoshitaka Ushiku, and Tatsuya Harada.
\newblock Learning from between-class examples for deep sound recognition.
\newblock \emph{CoRR}, abs/1711.10282, 2017.
\newblock URL \url{http://arxiv.org/abs/1711.10282}.

\bibitem[Tzanetakis and Cook(2002)]{tzanetakis2002musical}
George Tzanetakis and Perry Cook.
\newblock Musical genre classification of audio signals.
\newblock \emph{IEEE Transactions on speech and audio processing}, 10\penalty0
  (5):\penalty0 293--302, 2002.

\bibitem[Wikipedia(2021{\natexlab{a}})]{wiki:tango}
Wikipedia.
\newblock Tango.
\newblock \url{https://en.wikipedia.org/wiki/Tango}, 2021{\natexlab{a}}.
\newblock [Online; accessed 21-April-2023].

\bibitem[Wikipedia(2021{\natexlab{b}})]{wiki:tango-music}
Wikipedia.
\newblock Tango music.
\newblock \url{https://en.wikipedia.org/wiki/Tango_music}, 2021{\natexlab{b}}.
\newblock [Online; accessed 21-April-2023].

\bibitem[Yang et~al.(2022)Yang, Yu, Wang, Wang, Weng, Zou, and
  Yu]{yang2022diffsound}
Dongchao Yang, Jianwei Yu, Helin Wang, Wen Wang, Chao Weng, Yuexian Zou, and
  Dong Yu.
\newblock Diffsound: Discrete diffusion model for text-to-sound generation.
\newblock \emph{arXiv preprint arXiv:2207.09983}, 2022.

\end{thebibliography}



\end{document}